\documentclass[12pt]{article}
\usepackage{amssymb}
\usepackage{graphicx}
\usepackage{amsmath}

\begin{document}
\begin{center}
{\Large Extended gcd of quadratic integers \medskip}
\end{center}
Abdelwaheb Miled \qquad\qquad\qquad\qquad\qquad\qquad Ahmed Ourtani\\
Universit\'e de Sousse ISSAT Tunisia\qquad\qquad Facult\'e des Sciences Tunis Tunisia\\
abdelwahebmiled@gmail.com \qquad\qquad\qquad\qquad ahmedouertani@yahoo.fr
\\
\textbf{Abstract}{\small :}
{\small Computation of the  extended gcd of two quadratic integers. The ring of integers considered is principal but could be euclidean or not euclidean ring. This method rely on principal ideal  ring and  reduction of binary quadratic forms  .\bigskip }

\textbf{1-Notations and preliminaries}\\ 
If $d$ is a positive square free integer  , then $\mathbb{Q(}\sqrt{d}\mathbb{)}$
is a quadratic extension of $\mathbb{Q}$, the ring of algebraic integers of  $\mathbb{Q(}\sqrt{d}\mathbb{)}$ is $\mathbb{Z}[\theta ]=\left\{
a+b\theta \text{ },\text{ }a,b\in \mathbb{Z}\right\} ,$ where $\theta=\dfrac{1+\sqrt{d}}{2}$ if  $d \mod 4=1$ and $\theta=\sqrt{d}$ if $d \mod 4=2,3$.

The purpose of this paper is to compute the extended gcd of to quadratic integers in ring  $\mathbb{Z}[\theta ]$. We assume throughout that $\mathbb{Z}[\theta ]$ is principal ideal ring, but not necessarily an euclidean ring. \\
If $[a,b+c \theta]$ is the module $\{a x + (b+c \theta) y, x,y \in \mathbb{Z}\}$, it can be shown  \cite{Samuel}
that $I$ is an ideal of $\mathbb{Z}[\theta]$ if and only if $I=[a,b+c\theta]$; where $a,b,c \in \mathbb{Z}$ with $c|b,c|a$ and $ac|N(b+c\theta)$, ($N(b+c\theta)$ the norm of $b+c\theta$).
It is known \cite[Samuel] that any non-zero ideal of $\mathbb{Z}[\theta]$ is a free $\mathbb{Z}$-module of range $2$\\
If $I=\mathbb{Z\alpha }$ \ $\mathbb{+} \ \mathbb{Z}(\beta +\gamma \theta)$  is an ideal of $\mathbb{Z}[\theta]$, then  $I$ have a
generator for which the absolute value of the norm is equal to the norm $N(I) $ of $I$ (where $N(I)=\left| \alpha \times \gamma \right| $).
Conversely any $X$ in $I$ such that $|N(X)|=N(I)$ is a generator of  $I$.\\
\textbf{2-Algorithm of the extended gcd}\\
Let $I=[a+b\theta],J=[c+d \theta]$  non-zero ideals of $\mathbb{Z}[\theta]$, then $I+J=[m+n\theta]$ is an ideal, and $m+n\theta =\gcd( (a+b\theta),(c+d\theta))$.
This gcd is defined modulo a unity of $\mathbb{Q(}\sqrt{d}\mathbb{)}$. There exist $U,V\in \mathbb{Z}[\theta ]$ such that 
\begin{equation*}
U.(a+b\theta )+V.(c+d\theta )=m+n\theta
\end{equation*}
Since $I+J$ is maximal, it has a generator $\mathbb{Z}$-system of the form $\mathbb{Z\alpha +Z(}\beta +\gamma \theta )$
where $\gamma |\alpha $ and $\gamma |\beta $ and satisfy $| N(m\alpha +n(\beta +\gamma \theta ))| =| \alpha \gamma|$,
and $m,n\in \mathbb{Z}$ are solution of 
\begin{equation*}
\left| \frac{\alpha }{\gamma }x^{2}+(\frac{2\beta }{\gamma }+S)xy+\frac{\beta ^{2}+\beta \gamma S+\gamma ^{2}P}{\alpha \gamma }y^{2}\right| =1
\end{equation*}
To solve this equation we consider the binary  quadratic form 
\begin{equation*}
q(x,y)=\frac{\alpha }{\gamma }x^{2}+(\frac{2\beta }{\gamma }+S)xy+\frac{\beta ^{2}+\beta \gamma S+\gamma ^{2}P}{\alpha \gamma }y^{2}
\end{equation*}
If this form is positive definite, then \cite{Buell} it will be reduced to  one of  following  reduced forms 
\begin{eqnarray*}
&&[1,1,1],[1,0,1],[1,1,2],[1,0,2],[1,1,3] ,[1,1,5],[1,1,11],[1,1,17],[1,1,41]
\end{eqnarray*}
If the form is indefinite,  it will be reduced to either principal quadratic form $q_{p}(1,b,c,x,y)$ or $q_{p^{-}}(-1,b,-c,x,y)$. In the first case  $q_{p}(1,b,c,1,0)=1$, and in the second $q_{p^{-}}(-1,b,-c,1,0)=-1$.
\\
The main steps of this algorithm are :\\
Let $X=a_{1}+a_{2}\theta,Y=b_{1}+b_{2}\theta \in \mathbb{Z}[\theta]$,  relatively prime, and consider the ideal $(X\mathbb{Z}[\theta ]+Y\mathbb{Z}[\theta])$, it is generated by $X,X\theta,Y,Y\theta$.
\\
 Let $S=\theta+\bar{\theta},P=\theta \bar{\theta},d_{1}=gcd(a_{1},a_{1}+a_{2}S),d_{2}=gcd(b_{2},b_{1}+b_{2}S)$ :
\begin{center}
$\exists u_{1},v_{1} \in \mathbb{Z} \qquad u_{1}a_{2}+v_{1}(a_{1}+a_{2}S)=d_{1}$\\
$\exists  u_{2},v_{2} \in \mathbb{Z} \qquad u_{2}b_{2}+v_{2}(b_{1}+b_{2}S)=d_{2}$
\end{center}
If $d_{3}=gcd(d_{1},d_{2})$ then: $d_{3}=u_{3}d_{1}+v_{3}d_{2}, u_{3},v_{3} \in \mathbb{Z}$\\
Substitute  $\left( X,X\theta \right) $ and $\left( Y,Y\theta \right) $ by 
\begin{equation*}
\left( \left( \dfrac{a_{1}+a_{2}S}{d_{1}}-\dfrac{a_{2}}{d_{1}}\theta \right)
X\text{ },\text{ }(u_{1}+v_{1}\theta )X\right)\\
\left( \left( \dfrac{b_{1}+b_{2}S}{d_{2}}-\dfrac{b_{2}}{d_{2}}\theta \right)
Y\text{ },\text{ }(u_{2}+v_{2}\theta )Y\right)
\end{equation*}
Now, substitute $\left((u_{1}+v_{1}\theta )X)\text{ },\text{ }(u_{2}+v_{2}\theta )Y\right) $ by 
\begin{equation*}
\left( \left( \dfrac{d_{2}}{d_{3}}\right) (u_{1}+v_{1}\theta )X-\left( 
\dfrac{d_{1}}{d_{3}}\right) (u_{2}+v_{2}\theta )Y\text{ },\text{ }
u_{3}(u_{1}+v_{1}\theta )X+v_{3}(u_{2}+v_{2}\theta )Y\right)
\end{equation*}
Compute extended gcd of three integers : 
\begin{equation*}
\dfrac{a_{1}+a_{2}S}{d_{1}}+\dfrac{a_{2}^{2}}{d_{1}}P\text{ \ },\text{ \ }
\dfrac{b_{1}+b_{2}S}{d_{2}}-\dfrac{b_{2}^{2}}{d_{2}}P\text{ },\text{ \ \ }
\dfrac{d_{2}}{d_{3}}(u_{1}-v_{1}P)-\dfrac{d_{1}}{d_{3}}(u_{2}-v_{2}P)
\end{equation*}
This yields $\alpha ,\beta ,\gamma \in \mathbb{Z}$ such that 
\begin{equation*}
\alpha \left( \dfrac{a_{1}+a_{2}S}{d_{1}}-\dfrac{a_{2}}{d_{1}}\theta \right)
X+\beta \left( \dfrac{b_{1}+b_{2}S}{d_{2}}-\dfrac{b_{2}}{d_{2}}\theta
\right) Y+\gamma \left( \dfrac{d_{2}}{d_{3}}(u_{1}+v_{1}\theta )X-\dfrac{d_{1}}{d_{3}}(u_{2}+v_{2}\theta )Y\right) =1
\end{equation*}
i.e: 
\begin{equation*}
UX+VY=1
\end{equation*}
where 
\begin{eqnarray*}
U &=&\alpha \left( \dfrac{a_{1}+a_{2}S}{d_{1}}-\dfrac{a_{2}}{d_{1}}\theta
\right) +\gamma \left[ \dfrac{d_{2}}{d_{3}}(u_{1}+v_{1}\theta )\right] \\
V &=&\beta \left( \dfrac{b_{1}+b_{2}S}{d_{2}}-\dfrac{b_{2}}{d_{2}}\theta
\right) -\gamma \left[ \dfrac{d_{1}}{d_{3}}(u_{2}+v_{2}\theta )\right] \text{
\ \ \ \ \ \ \ \ \ \ \ \ \ \ \ }
\end{eqnarray*}
\textbf{Exemples}\\
1-The  ring $ \mathbb{Z}[\dfrac{1+\sqrt{-19}}{2}]$ is principal i but not euclidean \cite{Campoli}.\\
Let $X = -70+93 \theta, Y = -45 + 103 \theta \in \mathbb{Z}[\dfrac{1+\sqrt{-19}}{2}], \gcd(X,Y)=5 + 2 \theta$
\begin{center}
$5+2 \theta=(50+6 \theta)X+(-85+164 \theta)Y$
\end{center}
2- $X = 92+73 \theta,Y = 59+46 \theta \in \mathbb{Z}[\dfrac{1+\sqrt{13}}{2}], \gcd(X,Y)=1-\theta$:
\begin{center}
$1-\theta=(50+38 \theta)X+(-81-59 \theta)Y$
\end{center}
3-$X = 290 + 55 \theta,Y = 180 + 35 \theta \in \mathbb{Z}[\sqrt{14}],  \gcd(X,Y)=10 - 5 \theta$:
\begin{center}
$10-5 \theta=(3+3 \theta)X+(-4-5 \theta)Y$
\end{center}
4-$X = -70+93 \theta,Y = -40+61 \theta \in \mathbb{Z}[\sqrt{-2}], \gcd(X,Y)= 2+3 \theta$
\begin{center}
$2+3 \theta=(20+35 \theta)X+(-27-56 \theta)Y$
\end{center}

\bigskip
\hrule
\bigskip

\noindent 2000 {\it Mathematics Subject Classification}:
11R11.11R09,11R29
\noindent \emph{Keywords: }ring of integers- extended gcd of quadratic integers -ideal-binary quadratic form.
\end{document}